# Limit on the Addressability of Fault-Tolerant Nanowire Decoders

Yeow Meng Chee, *Senior Member*, *IEEE*, and Alan C.H. Ling

**Abstract**—Although prone to fabrication error, the nanowire crossbar is a promising candidate component for next-generation nanometer-scale circuits. In the nanowire crossbar architecture, nanowires are addressed by controlling voltages on the mesowires. For area efficiency, we are interested in the maximum number of nanowires $N(m,e)$ that can be addressed by $m$ mesowires in the face of up to $e$ fabrication errors. Asymptotically tight bounds on $N(m,e)$ are established in this paper. In particular, it is shown that $N(m,e) = \Theta(2^m/m^{e+1/2})$. Interesting observations are made on the equivalence between this problem and the problem of constructing optimal error-correcting and all unidirectional error-detecting (EC/AUED) codes, superimposed distance codes, pooling designs, and diffbounded set systems. Results in this paper also improve upon those in the EC/AUED code literature.

✦

## 1 INTRODUCTION

THE semiconductor industry today relies on photolithography techniques to transfer design patterns onto silicon wafers. Chips with 90-nm features are now in mass production, and NAND flash memories with feature size of 40 nm are about to debut. However, feature sizes need to get smaller to ensure the continuation of Moore's Law. Ultimately, the limits of photolithography, which industry consensus holds to be around 30 nm, will be reached. New methods for growing and assembling nanometer-scale wires (*nanowires*) are therefore pursued to extend Moore's Law beyond the limits of photolithography [1], [2], [3], [4], [5].

One promising technology in this direction is that of nanoarray architecture, notably, the *nanowire crossbar*. The small size and high density of these structures make them favorable candidates for future high-density interconnect, computation, and information storage devices [6], [7], [8], [9], [10], [11], [12], [13], [14], [15], [16], [17], [18]. Furthermore, the nanowire crossbar is the only nanoscale architectural component that has been fabricated to date [5], [15], [19].

A nanowire crossbar consists of two orthogonal sets of parallel nanowires separated by a molecular layer. Within this layer, molecules at the crosspoints of pairs of orthogonal nanowires can switch their conductivity under the application of large positive and negative electric fields [9], [20], [21], [22]. The conductivity at crosspoints can also be sensed without changing it by the application of smaller electric fields.

A fundamental challenge in crossbar architectures is providing a reliable means of controlling individual nanowires in each dimension. A circuit that provides this control is called a *nanowire decoder*. Current state of art requires that nanowire decoders be implemented in CMOS technologies. A nanowire decoder addresses the nanowires through an interface of a small set of *mesowires*—wires of mesoscale feature size (100-500 nm). Since mesowires are considerably larger and take up more area, the objective is to maximize the number of nanowires that can be addressed by a given set of mesowires. While it is obvious that $m$ mesowires can generate $2^m$ states, the characteristics of nanowire crossbars prevent us from achieving this limit. Furthermore, the process of nanowire crossbar fabrication is error prone. There is a need, therefore, to build in redundancy to tolerate such faults [23]. In a series of papers, Rachlin and Savage [24], [25] have established criteria and bounds on fault-tolerant nanowire decoders. However, their bounds become weaker when the ability to assemble nanowire crossbars become more deterministic (see Section 8.1 for a more precise statement and comparison).

In this paper, we give the first provably tight bound (up to a constant factor) on $N(m,e)$, the maximum number of nanowires that can be addressed by a nanowire decoder with $m$ mesowires in the face of up to $e$ fabrication errors. We achieve this by casting the problem in the language of extremal set systems and observing its equivalence to many other well-studied problems, including error-correcting and all unidirectional error-detecting (EC/AUED) codes, superimposed distance codes, pooling designs, and diffbounded set systems. Exact values of $N(m,e)$ are also obtained for some parameter sets $(m,e)$, including an infinite family based on Hadamard designs.

We note that in practice, the assembly of nanowire decoders is a random process [26], [27]. The mesowires that control a nanowire cannot be specified in advance. Hence, our bounds and exact values of $N(m,e)$ set upper limits on the number of independently addressable nanowires.

The results in this paper also improve upon results in the EC/AUED code literature.

- *Y.M. Chee is with the Division of Mathematical Sciences, School of Physical and Mathematical Sciences, Nanyang Technological University, 21 Nanyang Link, Singapore 637371. E-mail: ymchee@ntu.edu.sg.*
- *A.C.H. Ling is with the Department of Computer Science, University of Vermont, Burlington, VT 05405. E-mail: aling@emba.uvm.edu.*

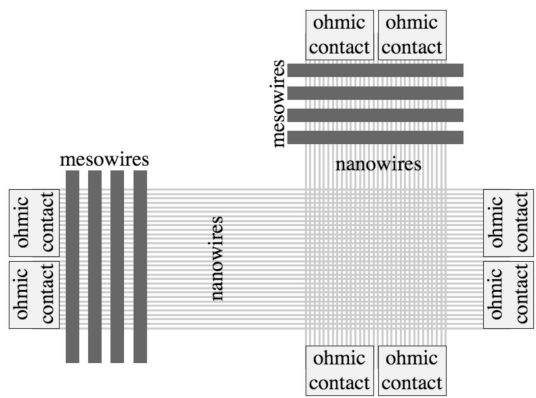

Fig. 1. The nanowire crossbar. Shown here with two orthogonal sets of parallel nanowires controlled by mesowires. Ohmic contacts are made at the ends of each set of parallel nanowires. Data is stored in the programmable molecules at the intersections of orthogonal nanowires.

## 2 THE NANOWIRE CROSSBAR

We begin by reviewing the nanowire crossbar. The model takes the form in Fig. 1.

Current methods of nanowire production results in two types of nanowires:

1. *Undifferentiated nanowires.* These nanowires are all identical. Methods for producing undifferentiated nanowires include *superlattice nanowire pattern transfer* (SNAP) [4], *porous membranes* [28], and *on-chip catalysts* [29].
2. *Differentiated nanowires.* To differentiate nanowires, dopant molecules are added to a gaseous mixture as they grow. As a result, nanowires are heavily and lightly doped over their lengths, depending on the exposure time (see Fig. 2). The two primary methods of doping are *axial (or modulation) doping* [30], [31], [32], [33], [34] and *radial doping* [35], [36].

Lithographically produced pairs of *ohmic contacts* are attached to both ends of a set of parallel nanowires. This allows a potential to be applied across all nanowires attached to a pair of ohmic contacts. Each pair of ohmic contacts can provide voltage control over sets of about 10 to 20 nanowires. Ohmic contacts can be reliably controlled using standard CMOS circuitry.

When a pair of ohmic contacts applies a potential across a set of parallel nanowires, two nanowires in the set carry the same current unless their resistances are different. Different resistances can be created on the nanowires via several methods, depending on whether differentiated or undifferentiated nanowires are used:

1. *Creating resistance on differentiated nanowires.* The resistance of a lightly doped silicon nanowire significantly increases in the presence of a sufficiently strong electric field. If a lithographically produced mesowire is laid down across a set of lightly doped nanowires, the nanowires only conduct when the mesowire is not producing an electric field. In this way, the mesowire forms a field-effect transistor (FET) with each nanowire [30]. If each mesowire forms an FET with only some of the nanowires, multiple mesowires can be used simultaneously to gain fine-grained control over nanowire resistances. For example, some nanowires may contain lightly doped regions under some mesowires and heavily doped regions under others (see Fig. 3). If a subset of the mesowires all produce an electric field simultaneously, all nanowires with a lightly doped region under any one of those mesowires will have a high resistance.
2. *Creating resistance on undifferentiated nanowires.* Contact between a mesowire and a nanowire is made by depositing impurities such as gold particles at the contacts [10] or depositing a high-K dielectric at the contacts. To create resistance between a mesowire and a nanowire, we prevent the deposition of such impurities at their contact.

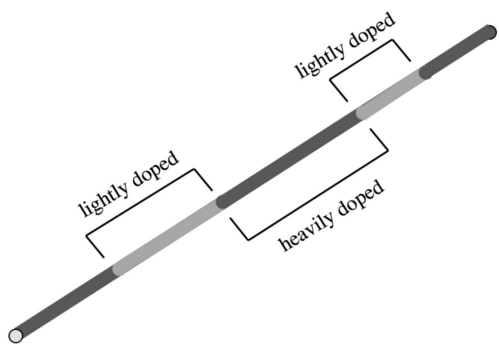

Fig. 2. Doped nanowire.

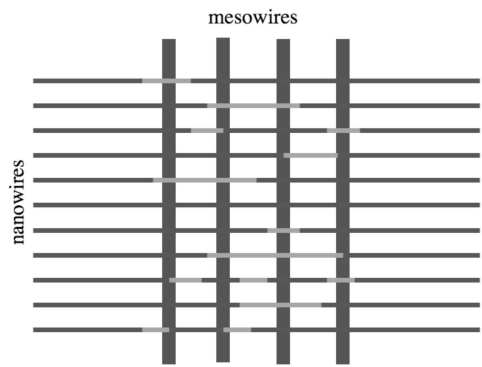

Fig. 3. Nanowires containing lightly doped and heavily doped regions.

## 3 NANOWIRE DECODERS

A voltage differential must be established across a pair of orthogonal nanowires in order to change or sense the state of molecules at their intersection. As discussed earlier, a pair of ohmic contacts provide the voltage control across a set of nanowires. Together with the mesowires across the set of nanowires, this is called a *simple nanowire decoder* [24]. A collection of multiple simple nanowire decoders controlling sets of nanowires in the same dimension is called a *composite nanowire decoder* [24] (see Fig. 4). It suffices to study simple nanowire decoders since in a composite nanowire decoder, the constituent simple nanowire decoders address disjoint sets of nanowires.

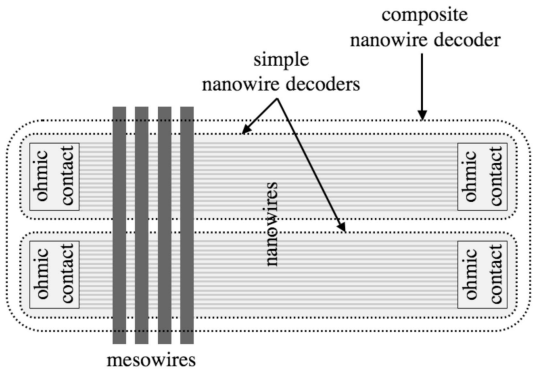

Fig. 4. Nanowire decoders.

More precisely, a simple nanowire decoder possesses the following properties:

1. $m$ mesowires control $n$ nanowires, which are tightly packed and aligned in the same direction but are not in electrical contact with each other.
2. A pair of ohmic contacts applies a voltage across all nanowires simultaneously. In the absence of mesowires, all nanowires conduct, effectively behaving like a single wire.
3. Each mesowire provides control over some subset of nanowires. A mesowire controls a nanowire if its resistance increases substantially when that mesowire carries a voltage.
4. When multiple mesowires carry a voltage, the resistance of a nanowire under the control of those mesowires is the sum of the resistance of that nanowire under the control of each mesowire.

If a nanowire has high resistance when a voltage is applied on a mesowire, the nanowire is said to be *controlled* by that mesowire. A configuration of voltage application on the set of all mesowires is called an *activation pattern*. A nanowire that has low resistance under some activation pattern is said to be *addressed*.

Let $S$ be the set of all nanowires of a simple nanowire decoder and $T \subseteq S$. Then, $T$ is said to be *addressable* if there exists some activation pattern under which all nanowires in $T$ are addressed and all nanowires not in $T$ are not addressed. In a simple nanowire decoder, only those nanowires $i \in S$ for which $\{i\}$ is addressable is useful, in order to change or sense the state of a single molecular element at the crosspoints. We call such nanowires *independently addressable* [24].

The process of fabricating nanowire crossbars is subject to error. A mesowire $i$ may end up with unpredictable control over a nanowire $j$. This is due to the error-prone process of doping nanowires and the deposition of impurities or high-K dielectric at the contacts. In such a situation, we want to avoid activating mesowire $i$ when addressing nanowire $j$ [24].

The problem of interest is in the determination of the limit of a fault-tolerant simple nanowire decoder: What is the maximum number of nanowires that can be independently addressed by a simple nanowire decoder (with $m$ mesowires) in the presence of up to $e$ errors?

This problem is made more precise in Section 5. We begin by reviewing some necessary mathematical background.

## 4 MATHEMATICAL PRELIMINARIES

### 4.1 Codes

The *Hamming n-space* is the set $\mathcal{H}(n) = \{0,1\}^n$ endowed with the *Hamming distance* $d_H$ defined as follows: for $\mathsf{u}, \mathsf{v} \in \mathcal{H}(n)$, $d_H(\mathsf{u}, \mathsf{v})$ is the number of positions where $\mathsf{u}$ and $\mathsf{v}$ differ. The *Hamming weight* of a vector $\mathsf{u} \in \mathcal{H}(n)$ is the number of positions in $\mathsf{u}$ with nonzero value and is denoted $w_H(\mathsf{u})$. The $i$th component of $\mathsf{u} \in \mathcal{H}(n)$ is denoted $\mathsf{u}_i$. The *support* of $\mathsf{u} \in \mathcal{H}(n)$, denoted supp($\mathsf{u}$), is the set $\{i : \mathsf{u}_i = 1\}$.

A subset of $\mathcal{H}(n)$ is called a *code of length* $n$. A *constant-weight code of length* $n$ and *weight* $w$ is any subset of $\mathcal{H}(n, w) = \{\mathsf{u} \in \mathcal{H}(n) : w_H(\mathsf{u}) = w\}$. The elements of a code are called *codewords*. Let $\mathcal{C} \subseteq \mathcal{H}(n)$ be a code. The *size* of $\mathcal{C}$ is $|\mathcal{C}|$, the number of codewords in the code. A code $\mathcal{C}$ is said to have *distance* $d$ if $d_H(\mathsf{u}, \mathsf{v}) \geq d$ for all distinct $\mathsf{u}, \mathsf{v} \in \mathcal{C}$. A code of distance at least $2e+1$ is said to be *e-error-correcting*.

The maximum size of a code of length $n$ and distance $d$ is denoted $A(n, d)$, while the maximum size of a constant-weight code of length $n$, weight $w$, and distance $d$ is denoted $A(n, d, w)$.

### 4.2 Set Systems

For $n$, a positive integer, let $[n]$ denote the set $\{1, 2, \ldots, n\}$. For a finite set $X$ and positive integer $k$, we define $2^X = \{A : A \subseteq X\}$ and $\binom{X}{k} = \{A \in 2^X : |A| = k\}$. We say that $A$ is a $k$-subset of $X$ if $A \in \binom{X}{k}$. Two sets $A$ and $B$ are *incomparable* if $A \nsubseteq B$ and $B \nsubseteq A$.

A *set system of order* $n$ is a pair $(X, \mathcal{A})$, where $X$ is a finite set of $n$ *points* and $\mathcal{A} \subseteq 2^X$. The elements of $\mathcal{A}$ are called *blocks*. A set system is said to be $k$-*uniform* if $\mathcal{A} \subseteq \binom{X}{k}$. The *size* of a set system $(X, \mathcal{A})$ is $|\mathcal{A}|$, the number of blocks. The *dual* of a set system $(X, \mathcal{A})$ is the set system $(Y, \mathcal{B})$, where $Y = \mathcal{A}$, and $\mathcal{B} = \cup_{x \in X}\{\{A \in \mathcal{A} : x \in A\}\}$.

An *antichain* is a set system $(X, \mathcal{A})$ such that for any distinct $A, B \in \mathcal{A}$, $A$ and $B$ are incomparable. Alternatively, an antichain may be defined as a set system $(X, \mathcal{A})$ in which $|A \setminus B| \geq 1$ for all distinct $A, B \in \mathcal{A}$.

Let $([n], \mathcal{A})$ be a set system. The *incidence vector* of a block $A \in \mathcal{A}$ is the vector $\iota(A) \in \mathcal{H}(n)$, where

$$\iota(A)_i = \begin{cases} 1, & \text{if } i \in A, \\ 0, & \text{otherwise.} \end{cases}$$

There is a natural correspondence between the Hamming $n$-space and the *complete* set system of order $n$ $([n], 2^{[n]})$: the positions of vectors in $\mathcal{H}(n)$ correspond to points in $[n]$, a vector $\mathsf{u} \in \mathcal{H}(n)$ corresponds to the block supp($\mathsf{u}$), and $d_H(\mathsf{u}, \mathsf{v}) = |\text{supp}(\mathsf{u}) \Delta \text{supp}(\mathsf{v})|$, where $\Delta$ denotes *symmetric difference*. From this, it follows that there is a bijection between the set of all codes of length $n$ and the set of all set systems of order $n$. Therefore, we may speak of the *set system of a code* or *the code of a set system*.

### 4.3 Designs

A $t$-*design*, or more specifically, a $t$-$(v, k, \lambda)$ *design*, is a $k$-uniform set system $(X, \mathcal{A})$ of order $v$ such that each $t$-subset of $X$ is contained in precisely $\lambda$ blocks of $\mathcal{A}$. For any $s \leq t$, a $t$-$(v, k, \lambda)$ design is also an $s$-$(v, k, \lambda_s)$ design, where $\lambda_s = \lambda\binom{v-s}{t-s}/\binom{k-s}{t-s}$. Hence, a $t$-$(v, k, \lambda)$ design has size $\lambda_0 = \lambda \binom{v}{t}/\binom{k}{t}$, and every point in a $t$-$(v,k,\lambda)$ design is contained in exactly $\lambda_1 = \lambda \binom{v-1}{t-1}/\binom{k-1}{t-1}$ blocks.

A $t$-$(v,k,1)$ design is called a *Steiner system* and is often denoted $S(t,k,v)$.

A $2$-$(v,k,\lambda)$ design is also commonly called a $(v,k,\lambda)$-*balanced incomplete block design* (BIBD). A $(v,k,\lambda)$-BIBD $(X,\mathcal{A})$ is *symmetric* if $|\mathcal{A}| = v$. The dual of a symmetric $(v,k,\lambda)$-BIBD is again a $(v,k,\lambda)$-BIBD, so that every two blocks in a symmetric $(v,k,\lambda)$-BIBD intersect in exactly $\lambda$ points.

A *Hadamard matrix* $H$ of order $n$ is an $n \times n\{\pm 1\}$-matrix such that $HH^T = nI$. A necessary condition for a Hadamard matrix of order $n$ to exist is $n = 1, 2$, or $n \equiv 0 \pmod 4$. The existence of Hadamard matrices has not been completely settled and constitutes a major problem in combinatorics. Nevertheless, much is known concerning their existence.

A symmetric $(4n+3, 2n+1, n)$-BIBD has been called a *Hadamard design*: it exists if and only if a Hadamard matrix of order $4(n+1)$ exists.

An $(r,\lambda)$-*design* is a set system $(X,\mathcal{A})$ where every point of $X$ is contained in exactly $r$ blocks and every 2-subset of $X$ is contained in exactly $\lambda$ blocks.

A $(1,q)$-*pooling design* (called $(1,q)$-*solution* by Balding and Torney [37]) is a set system $(X,\mathcal{A})$ whose dual $(Y,\mathcal{B})$ satisfies $|B_1 \setminus B_2| > q$ for all distinct $B_1, B_2 \in \mathcal{B}$. $(1,q)$-pooling designs are used for group testing one defective element in the presence of up to $q$ errors [37], [38].

The reader is referred to [39], [40], [41], [42], [43], [44], [45], and [46] for more information on the designs described in this section.

## 5 MATHEMATICAL MODEL

### 5.1 A Model for Simple Nanowire Decoders

Rachlin and Savage [24] model a simple nanowire decoder using codes. We model a simple nanowire decoder using set systems here. Although equivalent to the coding-theoretic model of Rachlin and Savage through the correspondence between codes and set systems described earlier, we find that the set system formulation is more natural, the proofs based on set systems are often simpler, and existing results on extremal set systems can be brought to bear. Now, we describe a mathematical model of a simple nanowire decoder in terms of set systems.

A simple nanowire decoder with $m$ mesowires and $n$ nanowires is denoted $(m,n)$-$SND$. For an $(m,n)$-SND $\mathcal{D}$, let $[m]$ denote the set of all mesowires and $[n]$ denote the set of all nanowires. For each nanowire $i \in [n]$, let $A_i$ be the set of all mesowires controlling $i$ and $\mathcal{A} = \{A_i : i \in [n]\}$. Then, $\mathcal{D}$ is completely specified by $([m], \mathcal{A})$. Hence, we represent an $(m,n)$-SND by a set system of order $m$ and size $n$. Note that we may identify the nanowire $i$ with the block $A_i \in \mathcal{A}$. An activation pattern is simply a set $V \subseteq [m]$ such that for $i \in [m]$, we have $i \in V$ if and only if mesowire $i$ carries a voltage. Hence, the following holds:

1. A nanowire $A \in \mathcal{A}$ is *addressed* under an activation pattern $V$ if $A \cap V = \emptyset$.
2. A nanowire $A \in \mathcal{A}$ is *independently addressable* if there exists an activation pattern $V$ such that $A \cap V = \emptyset$ and $B \cap V \neq \emptyset$ for all $B \in \mathcal{A}$ and $B \neq A$.

We assume throughout this paper that the simple nanowire decoders we consider contain only nanowires that are independently addressable. Those nanowires that are not independently addressable can be thrown out since they do not perform any useful function.

### 5.2 A Model for Fault-Tolerant Simple Nanowire Decoders

Consider an $(m,n)$-SND $\mathcal{D} = ([m], \mathcal{A})$. An *error* during the fabrication of $\mathcal{D}$ can be modeled as follows: The control of a nanowire $A \in \mathcal{A}$ by mesowire $j \in [m]$ becomes unpredictable. This unpredictability renders mesowire $j$ useless as a control over nanowire $A$. We can never rely on $j$ to control $A$. This is equivalent to deleting the element $j$ from $A$. We call an $(m,n)$-SND *fault-tolerant* if all the nanowires remain independently addressable in the presence of up to $e$ errors and denote such a decoder by $(m,n,e)$-$FTSND$. Note that an $(m,n)$-SND is equivalent to an $(m,n,0)$-FTSND.

For any given $m$ and $e$, the maximum $n$ such that there exists an $(m,n,e)$-FTSND is denoted $N(m,e)$.

## 6 THE LIMIT OF NANOWIRE DECODERS WITHOUT ERRORS

We begin by characterizing those nanowires that are independently addressable by $\mathcal{D}$.

**Proposition 1.** *Let $\mathcal{D} = ([m], \mathcal{A})$ be an $(m,n)$-SND. A nanowire $A \in \mathcal{A}$ is independently addressable if and only if $B \notin \mathcal{A}$ for all $B \subsetneq A$.*

**Proof.** Suppose $A \in \mathcal{A}$ is independently addressable and suppose that $B \in \mathcal{A}$ for some $B \subsetneq A$. Then, there exists an activation pattern $V$ such that $A \cap V = \emptyset$. But this implies $B \cap V = \emptyset$, which contradicts the assumption that $A$ is not independently addressable. Therefore, $B \notin \mathcal{A}$ for all $B \subsetneq A$.

Next, suppose that $B \notin \mathcal{A}$ for all $B \subsetneq A$. Let $V = [m] \setminus A$. Then, $A$ is independently addressable since $A \cap V = \emptyset$ and $B \cap V \neq \emptyset$ for all $B \in \mathcal{A}$ and $B \neq A$.

This completes the proof. □

**Corollary 1.** *$\mathcal{D}$ is an $(m,n)$-SND if and only if it is an antichain of order $m$ and size $n$.*

The problem of determining the maximum size of an antichain of order $m$ is the prototype problem in the combinatorics of finite sets and has been solved by Sperner [47].

**Theorem 1 (Sperner's Theorem).** *The maximum size of an antichain of order $m$ is $\binom{m}{\lfloor m/2 \rfloor}$. The set system $([m], \binom{[m]}{\lfloor m/2 \rfloor})$ achieves the maximum size.*

**Corollary 2.** $N(m,0) = \binom{m}{\lfloor m/2 \rfloor}$.

There are many proofs of Sperner's Theorem (see [48] and [49]). Rachlin and Savage [24] were apparently unaware of Sperner's Theorem and provided a proof of Sperner's Theorem using exactly the same *shifting technique* as in Sperner's original proof [47].

## 7 THE LIMIT OF FAULT-TOLERANT NANOWIRE DECODERS

We begin with a characterization of fault-tolerant simple nanowire decoders.

**Proposition 2.** *Let $\mathcal{D} = ([m], \mathcal{A})$ be a set system of order $m$ and size $n$. Then, $\mathcal{D}$ is an $(m, n, e)$-FTSND if and only if $|A \setminus B| \geq e + 1$ for all distinct $A, B \in \mathcal{A}$.*

**Proof.** Suppose $\mathcal{D}$ can tolerate up to $e$ errors and suppose that $|A \setminus B| \leq e$ for some distinct $A, B \in \mathcal{A}$. Consider the (up to $e$) errors that result in the deletion of elements in $A \setminus B$ from $A$. Then, the resulting set system is not an antichain, and hence, not all nanowires are independently addressable. This contradicts the assumption that $([m], \mathcal{A})$ can tolerate up to $e$ errors. Hence, $|A \setminus B| \geq e + 1$.

Now, assume that $|A \setminus B| \geq e + 1$ for all distinct $A, B \in \mathcal{A}$. Deletion of up to $e$ elements from/to the blocks of $\mathcal{A}$ will still result in a set system $([m], \mathcal{A}')$ in which $|A \setminus B| \geq 1$ for all distinct $A, B \in \mathcal{A}'$. But this is equivalent to saying that $([m], \mathcal{A}')$ is an antichain. Consequently, all the nanowires remain independently addressable. Hence, $([m], \mathcal{A})$ can tolerate up to $e$ errors. □

A set system $([m], \mathcal{A})$ with the property

$$|A \setminus B| \geq e + 1, \quad \text{for all distinct } A, B \in \mathcal{A}, \tag{1}$$

has been called $(e+1)$-*diffbounded* by Katona [50]. Such set systems have been studied in attempts to further generalize Sperner's Theorem. They were also studied earlier under the following guises:

1. *By Pradhan in 1980* [51]. A code $C \subseteq \mathcal{H}(m)$ is ($e$-EC/AUED) if and only if the set system $([m], \mathcal{A})$ of $\mathcal{C}$ possesses the property (1).
2. *By D'yachkov et al. in 1989* [52]. A code $C \subseteq \mathcal{H}(m)$ is a *superimposed code of distance $e + 1$* if and only if the set system $([m], \mathcal{A})$ of $\mathcal{C}$ possesses the property (1).
3. *By Balding and Torney in 1996* [37]. A set system $(Y, \mathcal{B})$ of size $m$ is a $(1, e)$-*pooling design* if and only if its dual $([m], \mathcal{A})$ possesses the property (1).

Other related combinatorial objects include *disjunct matrices* [53] and *generalized cover-free families* [54]. We summarize these observations below.

**Proposition 3.** *The following are all equivalent:*

1. *an $(m, n, e)$-FTSND,*
2. *an $(e + 1)$-diffbounded set system of order $m$ and size $n$,*
3. *a $(1, e)$-pooling design of order $n$ and size $m$,*
4. *a superimposed code of length $m$, distance $e + 1$, and size $n$, and*
5. *an $e$-EC/AUED code of length $m$ and size $n$.*

The results on diffbounded set systems, $(1, e)$-pooling designs, superimposed codes, and EC/AUED codes can now be brought to bear. In particular, we have the following upper bound from the treatment of Balding and Torney [37] on $(1, e)$-pooling designs.

**Theorem 2 (Balding and Torney [37]).**

$$N(m, e) \leq \frac{1}{K_e} \binom{m}{\lfloor m/2 \rfloor},$$

in which $K_0 = 1$ and

$$K_e = \begin{cases} \sum_{s=0}^{e/2} \binom{\lfloor m/2 \rfloor}{s} \binom{\lceil m/2 \rceil}{s}, & \text{if } e \text{ is even,} \\ K_{e-1} + \frac{1}{T} \binom{\lfloor m/2 \rfloor}{(e+1)/2} \binom{\lceil m/2 \rceil}{(e+1)/2}, & \text{if } e \text{ is odd,} \end{cases}$$

where

$$T = \left\lfloor \frac{2}{e+1} \left\lfloor \frac{m}{2} \right\rfloor \right\rfloor.$$

**Corollary 3 (Balding and Torney [37]).** *If there exists an $S(\lfloor m/2 \rfloor - 1, \lfloor m/2 \rfloor, m)$, then*

$$N(m, 1) = \binom{m}{\lfloor m/2 \rfloor - 1} / \lfloor m/2 \rfloor.$$

Corollary 3 together with the current knowledge on the existence of Steiner systems (see [40]) then gives the following result.

**Corollary 4.** *We have the following:*

1. $N(4, 1) = 2$.
2. $N(7, 1) = 7$.
3. $N(8, 1) = 14$.
4. $N(11, 1) = 66$.
5. $N(12, 1) = 132$.

**Proof.**

1. The proof follows from the trivial $S(1, 2, 4)$.
2. The proof follows from the existence of $S(2, 3, 7)$.
3. The proof follows from the existence of $S(3, 4, 8)$.
4. The proof follows from the existence of $S(4, 5, 11)$.
5. The proof follows from the existence of $S(5, 6, 12)$. □

There are no $S(t, k, v)$ known for any $t \geq 6$ [40].

Proposition 3 also implies that the code of an $(m, n, e)$-FTSND is a code of length $m$, distance $2(e + 1)$, and size $n$. Hence, we have the following result.

**Proposition 4.** $N(m, e) \leq A(m, 2(e + 1))$.

The Plotkin bound [55] on the size of codes then implies the following:

**Corollary 5 (Plotkin bound).**

$$N(m, e) \leq \begin{cases} 2 \left\lfloor \frac{2(e+1)}{4(e+1)-m} \right\rfloor, & \text{if } m < 4(e+1), \\ 8(e+1), & \text{if } m = 4(e+1). \end{cases}$$

### 7.1 Further Exact Values of $N(m, e)$

In this section, we establish some more exact values of $N(m, e)$.

**Proposition 5.** $N(m, e) = 1$ if $m \leq 2e + 1$.

**Proof.** If $N(m,e) > 1$, then there are two distinct blocks $A$ and $B$ such that $|A \setminus B| \geq e+1$ and $|B \setminus A| \geq e+1$. Hence, $m \geq |A \cup B| \geq |A \setminus B| + |B \setminus A| \geq 2e+2$. □

**Proposition 6.** $N(m,e) = 2$ if $2e+2 \leq m \leq 3e+2$.

**Proof.** The Plotkin bound gives $N(m,e) \leq 2$ when $m \leq 3e+2$. When $m \geq 2(e+1)$, taking two disjoint blocks, each containing $e+1$ elements gives an $(e+1)$-diffbounded set system. The proposition then follows. □

**Proposition 7.** *If there exists a Hadamard matrix of order $4(e+1)$, then $N(4e+2, e) = 2e+2$.*

**Proof.** Suppose there exists a Hadamard matrix of order $4(e+1)$. Then, there exists a Hadamard $(4e+3, 2e+1, e)$-BIBD $(X, \mathcal{A})$. This design is symmetric so that any two blocks intersect in exactly $e$ points. Hence, $(X, \mathcal{A})$ is $(e+1)$-diffbounded.

Now, pick any $x \in X$ and consider $\mathcal{B} = \{A \in \mathcal{A} : x \notin A\}$. Then, $(X \setminus \{x\}, \mathcal{B})$ is a set system of order $4e+2$ and size $2e+2$ since the number of blocks in $\mathcal{A}$ containing $x$ is $2e+1$. Furthermore, $(X, \mathcal{B})$ is still $(e+1)$-diffbounded since $\mathcal{B} \subseteq \mathcal{A}$. This shows that $N(4e+2, e) \geq 2e+2$. The Plotkin bound gives $N(4e+2, e) \leq 2e+2$. Hence, we conclude that $N(4e+2, e) = 2e+2$. □

**Proposition 8.** *If there exists an $(r, \lambda)$-design of order $n$ and size $m$, then $N(m, r - \lambda - 1) \geq n$.*

**Proof.** Let $(X, \mathcal{A})$ be an $(r, \lambda)$-design of order $n$ and size $m$. Then, its dual $(Y, \mathcal{B})$ is an $r$-uniform set system of order $m$ and size $n$ such that every two blocks intersect in exactly $\lambda$ points. Hence, $(Y, \mathcal{B})$ is $(r - \lambda)$-diffbounded, and the proposition follows. □

**Corollary 6.** $N(6, 1) = 4$, and $N(20, 5) = 6$.

**Proof.** Apply Proposition 8 on the complete set system $([n], \binom{[n]}{k})$, which is a $(\binom{n-1}{k-1}, \binom{n-2}{k-2})$-design of order $n$ and size $\binom{n}{k}$, to obtain

$$N\left(\binom{n}{k}, \binom{n-2}{k-1} - 1\right) \geq n.$$

When $(n,k) = (4,2)$ and $(6,3)$, this gives $N(6,1) \geq 4$ and $N(20,5) \geq 6$, respectively. The matching upper bounds are given by the Plotkin bound. □

### 7.2 Bounds and Asymptotics

The following bounds on the size of $s$-diffbounded set systems were established by Katona [50].

**Theorem 3 (Katona [50]).** *If $(X, \mathcal{A})$ is an $s$-diffbounded set system of order $m$ and maximum size, then*

$$\alpha(s) \frac{2^m}{m^{s-1/2}} \leq |\mathcal{A}| \leq \beta(s) \frac{2^m}{m^{s-1/2}},$$

*where*

$$\alpha(s) = \sqrt{\frac{2}{\pi}} \frac{1}{2^s} - o(m),$$

$$\beta(s) = \sqrt{\frac{2}{\pi}} 2^{s-1}(s-1)! + o(m).$$

**Corollary 7.** *For any fixed $e$, we have*

$$N(m, e) = \Theta\left(\frac{2^m}{m^{e+1/2}}\right).$$

Corollary 7 therefore determines the limit of a fault-tolerant simple nanowire decoder up to a constant factor. Next, we establish an explicit lower bound on $N(m, e)$.

**Proposition 9.** $N(m, e) \geq \max_w A(m, 2(e+1), w)$.

**Proof.** Any constant-weight code of length $m$, weight $w$, and distance $2(e+1)$ is $(e+1)$-diffbounded. □

Levenshtein [56] generalized the Gilbert-Varshamov bound for general codes to constant-weight codes. We give its proof here as it provides a recipe (though not an efficient one) for the construction of codes achieving the bound.

**Theorem 4 (Levenshtein bound).**

$$A(n, 2d, w) \geq \frac{\binom{n}{w}}{\sum_{i=0}^{d-1} \binom{w}{i}\binom{n-w}{i}}.$$

**Proof.** The proof is via a greedy construction algorithm. We start with the space $S \subseteq \mathcal{H}(n)$ of all vectors of weight $w$ and the empty code $\mathcal{C} \subseteq S$. At each step, we take an arbitrary vector $\mathbf{u}$ of weight $w$ from $S$, adjoin it to $\mathcal{C}$, and remove from $S$ the vectors of weight $w$ contained in the Hamming sphere of radius $2d-1$ around $\mathbf{u}$. Repeat until $S$ is empty.

The bound then follows by noting that the initial size of $S$ is $\binom{n}{w}$ and that the number of binary vectors of weight $w$ contained in the Hamming sphere of radius $2d-1$ is $\sum_{i=0}^{d-1} \binom{w}{i}\binom{n-w}{i}$. □

**Corollary 8.**

$$N(m, e) \geq \frac{\binom{m}{\lfloor m/2 \rfloor}}{\sum_{i=0}^{e} \binom{\lfloor m/2 \rfloor}{i}\binom{\lceil m/2 \rceil}{i}}.$$

Though simple, Proposition 9 improves on the size of all $e$-EC/AUED codes known (and hence also $(m, n, e)$-FTSND and $(e+1)$-diffbounded set systems). A performance comparison is provided in the next section.

## 8 COMPARISONS AND COMPUTATIONAL RESULTS

### 8.1 Asymptotics

Suppose an $(m, n)$-SND $([m], \mathcal{A})$ is constructed by a random process as follows: For each nanowire $i$ and mesowire $j$, $j$ controls $i$ with probability $p$. An $(m, n)$-SND so constructed is called an *ideal randomized contact decoder*. Rachlin and Savage [24, Theorem 4.1] proved the following result.

**Theorem 5 (Rachlin and Savage [24]).** *All $n$ nanowires of an ideal randomized contact decoder $(m, n)$-SND are addressable in the presence of $e$ errors, with probability at least $1 - \epsilon$, if*

$$m \geq \frac{\left(e + \sqrt{e^2 + 4\ln(N^2/\epsilon)}\right)^2}{4p(1-p)}.$$

TABLE 1
Number of Information Bits $k$ in an $e$-EC/AUED Code of Length $m$

| $e$ | $m$ | $k$ (this paper) | best $k$ previously known | authority |
|---|---|---|---|---|
| 1 | 15 | **9** | 8 | [57], [60] |
| 1 | 22 | **15** | 14 | [57], [60] |
| 2 | 21 | **10** | 8 | [57], [60] |
| 2 | 23 | **11** | 10 | [57] |
| 3 | 27 | **12** | 8 | [57], [60] |

Corollary 7 shows that if we have just $m = 4\ln(n^2) \leq (e + \sqrt{e^2 + 4\ln(n^2/\epsilon)})^2/4p(1-p)$ mesowires, then we can address $\Theta(n^\alpha/polylog(n))$ nanowires, with probability one, where $\alpha = 8\ln 2 \approx 5.545$. This is much larger than the $n$ nanowires guaranteed by Theorem 5. Of course, this assumes that we can manufacture nanowire decoders deterministically.

### 8.2 Exact Values

Given an $e$-EC/AUED code of length $m$ and size $n$, the number of information bits is $k = \lfloor \log_2 n \rfloor$. The results in this paper give improvements over some of the best known $e$-EC/AUED codes.

The best 1-EC/AUED codes are provided in the table of Laih and Yang [57]. Our improvements here are shown in Table 1.

Improvements over the best 2-EC/AUED codes currently known are provided in Table 2.

In Table 3, we provide the current state of knowledge on $N(m, e)$, for $0 \leq e < m \leq 16$. Where exact values of $N(m, e)$ are not known, upper and lower bounds are provided. The lower bounds are based on Proposition 9 and the upper bounds are based on Proposition 4 (tables of the values of $A(n, d)$ and $A(n, d, w)$ can be found in [58] and [59], respectively). Exact values are obtained via exhaustive search.

## 9 CONCLUSION

In this paper, we study a particular limit of fault-tolerant nanowire decoders: how many nanowires can be addressed with $m$ mesowires in the face of $e$ fabrication errors? By casting this problem in the context of set systems, we observe that it is equivalent to many problems in coding theory and combinatorics that have been previously investigated. This observation allows us to establish general bounds and asymptotically tight bounds on the limit of fault-tolerant nanowire decoders. The results obtained in this paper also improve existing ones on EC/AUED codes.

TABLE 2
2-EC/AUED Codes of Length $m$ with $k$ Information Bits, $k \leq 14$

| $k$ | $m$ in [61] | $m$ in [62] | $m$ in [63] | $m$ (this paper) |
|---|---|---|---|---|
| 4 | 14 | 20 | 14 | **12** |
| 5 | 16 | 21 | 18 | **14** |
| 6 | 19 | 22 | 20 | **15** |
| 7 | 21 | 23 | 22 | **17** |
| 8 | 23 | 26 | 24 | **18** |
| 9 | 24 | 27 | 26 | **20** |
| 10 | 25 | 29 | 28 | **21** |
| 11 | 26 | 31 | 30 | **22** |
| 12 | 27 | 32 | 34 | **24** |
| 13 | 30 | 34 | 36 | **25** |
| 14 | 31 | 35 | 38 | **27** |

TABLE 3
The Value of $N(m, e)$, $0 \leq e < m \leq 16$

| $m$ \ $e$ | 0 | 1 | 2 | 3 | 4 | 5 | 6 | 7 |
|---|---|---|---|---|---|---|---|---|
| 1 |  |  |  |  |  |  |  |  |
| 2 | 2 |  |  |  |  |  |  |  |
| 3 | 3 |  |  |  |  |  |  |  |
| 4 | 6 | 2 |  |  |  |  |  |  |
| 5 | 10 | 2 |  |  |  |  |  |  |
| 6 | 20 | 4 | 2 |  |  |  |  |  |
| 7 | 35 | 7 | 2 |  |  |  |  |  |
| 8 | 70 | 14 | 2 | 2 |  |  |  |  |
| 9 | 126 | 18 | 3 | 2 |  |  |  |  |
| 10 | 252 | 36 | 6 | 2 | 2 |  |  |  |
| 11 | 462 | 66 | 11 | 2 | 2 |  |  |  |
| 12 | 924 | 132 | 22 | 4 | 2 | 2 |  |  |
| 13 | 1716 | 166–256 | 26–32 | 4 | 2 | 2 |  |  |
| 14 | 3432 | 325–512 | 42–64 | 8 | 2 | 2 | 2 |  |
| 15 | 6435 | 585–1024 | 70–128 | 15–16 | 3 | 2 | 2 |  |
| 16 | 12870 | 1170–2048 | 120–256 | 30–32 | 4 | 2 | 2 | 2 |

Unfilled entries are taken to be one.

## ACKNOWLEDGMENTS

The research of Yeow Meng Chee is supported in part by the the National Research Foundation of Singapore under Research Grant NRF-CRP2-2007-03, the Singapore Ministry of Education under Research Grant T206B2204, and Nanyang Technological University under Research Grant M58110040.

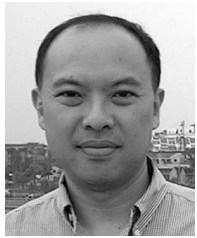

**Yeow Meng Chee** received the BMath degree in computer science and combinatorics and optimization and the MMath and PhD degrees in computer science from the University of Waterloo, Ontario, Canada, in 1988, 1989, and 1996, respectively. He is currently an associate professor in the Division of Mathematical Sciences, School of Physical and Mathematical Sciences, Nanyang Technological University, Singapore. Prior to this, he had been the program director of interactive digital media R&D in the Media Development Authority of Singapore, a postdoctoral fellow at the University of Waterloo and IBM's Zürich Research Laboratory, the general manager of the Singapore Computer Emergency Response Team, and the deputy director of strategic programs at the Infocomm Development Authority, Singapore. His research interest lies in the interplay between combinatorics and computer science/engineering, particularly combinatorial design theory, coding theory, extremal set systems, and electronic design automation. He is a senior member of the IEEE.

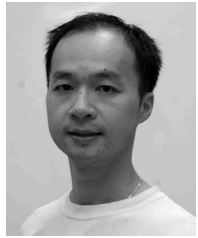

**Alan C.H. Ling** received the BMath, MMath, and PhD degrees in combinatorics and optimization from the University of Waterloo, Waterloo, Ontario, Canada, in 1994, 1995, and 1996, respectively. He worked at the Bank of Montreal, Montreal, Québec, and Michigan Technological University, Houghton, prior to his present position as an assistant professor of computer science at the University of Vermont, Burlington. His research interests concern combinatorial designs, codes, and applications in computer science.

▷ **For more information on this or any other computing topic, please visit our Digital Library at** www.computer.org/publications/dlib.